\documentclass[twocolumn]{article}
\usepackage{graphicx} % Required for inserting images
\usepackage{geometry}
\usepackage{booktabs}
\usepackage[table,xcdraw]{xcolor}
\title{Stock Market Sentiment Classification and Backtesting via Fine-tuned BERT }
\author{Jiashu Lou}
\date{}

\begin{document}

\maketitle

\begin{abstract}
    With the rapid development of big data and computing devices, low-latency automatic trading platforms based on real-time information acquisition have become the main components of the stock trading market, so the topic of quantitative trading has received widespread attention. And for non-strongly efficient trading markets, human emotions and expectations always dominate market trends and trading decisions. Therefore, this paper starts from the theory of emotion, taking East Money as an example, crawling user comment titles data from its corresponding stock bar and performing data cleaning. Subsequently, a natural language processing model BERT was constructed, and the BERT model was fine-tuned using existing annotated data sets. The experimental results show that the fine-tuned model has different degrees of performance improvement compared to the original model and the baseline model. Subsequently, based on the above model, the user comment data crawled is labeled with emotional polarity, and the obtained label information is combined with the Alpha191 model to participate in regression, and significant regression results are obtained. Subsequently, the regression model is used to predict the average price change for the next five days, and use it as a signal to guide automatic trading. The experimental results show that the incorporation of emotional factors increased the return rate by 73.8\% compared to the baseline during the trading period, and by 32.41\% compared to the original alpha191 model. Finally, we discuss the advantages and disadvantages of incorporating emotional factors into quantitative trading, and give possible directions for further research in the future.
\end{abstract}

\section{Introduction}
Emotion is a manifestation of internal thoughts that is intrinsic to human beings. In the field of artificial intelligence, understanding the emotional aspects of human language is essential. Emotion Recognition in Conversations (ERC) is a novel and emerging area of natural language processing (NLP) that aims to extract the sentiments expressed in conversational data from various sources such as Facebook, Twitter, and Weibo. By performing emotion recognition, ERC can identify the prevailing opinions and attitudes of the public on various topics and issues. Emotions are often associated with subjective feelings or environmental influences that are consciously perceived. Hence, emotions such as happiness, sadness, fear, anger, and surprise are derived from personal experiences and interactions with the external world. These emotions can affect an individual’s behavior and decision-making in a short span of time, as seen in financial markets where extreme emotions (such as euphoria or rage) can lead to irrational trading decisions, resulting in abnormal price fluctuations of assets.\par

Emotion analysis is a rapidly advancing field in machine learning today. Traditional sentiment classification models are based on the naive Bayes model (Murphy, 2006), which assigns the class with the highest posterior probability to a given sentence. With the advent of deep learning, neural networks have been increasingly applied to sentiment classification. The LSTM model, which has the ability to capture long-term and short-term dependencies in sequential data such as sentences, has a unique advantage in this task and is widely adopted in various sentiment classification models. The current state-of-the-art sentiment classification model is BERT (Bidirectional Encoder Representations from Transformers), which uses masked language modeling to pre-train bidirectional Transformers to generate deep bidirectional representations of language. It can better extract feature information from sentences and can be easily fine-tuned for downstream tasks.\par

In the financial domain, for a market that is not strongly efficient, traders’ emotions often affect the trading situation, eventually forming a consensus expectation that is reflected in the prices of the underlying assets. Therefore, monitoring and analyzing the public sentiment related to trading assets on forums and social media platforms is crucial for analyzing and predicting the prices of the assets. With the prevalence of online forums and social media, many people tend to express their opinions on stock price trends on the Internet (Duan et al., 2009; Kim and Kim, 2014). Previous studies have found a strong correlation between activities on stock discussion platforms and abnormal stock prices and trading volumes (Duan and Zeng, 2013; Tumarkin and Whitelaw, 2001). Based on the bounded rationality hypothesis in behavioral finance, individuals’ investment goals and decisions are not always consistent but rather exhibit some degree of confusion and inconsistency due to their cognitive levels and life experiences. Cognitive biases can easily influence the identification and resolution of problems, leading to suboptimal decision-making under the influence of emotions.\par

Automated trading systems are decision-making systems that rely on massive information. The reliability and performance of such systems depend largely on the efficiency of big data analysis and modeling. Predicting price trends in stock, commodity, and other derivative markets is a challenging task that attracts great interest from researchers and investors. It is influenced by numerous factors ranging from macroeconomics to participant emotions. Generally, there are two types of automated trading systems: fundamental analysis and technical analysis. Fundamental analysis focuses on the intrinsic value of company stocks (Lev and Thiagarajan, 1993), taking into account the past performance, future expectations, and political and economic environment of a specific company. Technical analysis, on the other hand, often ignores fundamental information and instead relies solely on past market data, especially price and volume data. Technical analysis assumes that all the information about the underlying asset is reflected in past prices and uses mathematical and statistical methods to discover useful factors and provide buy or sell signals based on combinations of these factors. As a result, it can quickly and sensitively detect moments of market price mismatches, enabling automatic arbitrage.\par

However, traditional automated trading systems often rely on market data such as prices and neglect alternative data sources such as forum posts, news texts, etc. Therefore, they lack sensitivity to sudden market sentiment shifts and retail investor speculation on individual stocks. Hence, we argue that using text-based sentiment analysis to uncover effective signal factors and develop automated trading systems is a promising and valuable research direction.\par

\section{Related works}
Since the beginning of the 21st century, the rapid development of computer technology, especially cloud computing, big data, and machine learning, has made it possible to acquire and analyze massive streams of real-time information. This has led both private and public fund companies to extensively utilize automated program trading for profit, engaging in rapid arbitrage of market price mismatches at the tick level. With the continuous advancement of machine learning and deep learning technologies, sentiment analysis in stock markets has gradually shifted from qualitative to quantitative, and from simple models to complex models. The following provides an overview of the domestic and international research progress in text sentiment analysis and its application in quantitative investment.

Text sentiment analysis was initially based on traditional classification models such as the Naive Bayes model. Modern baseline models were proposed by Bo Pang (2002) and others, who found that standard machine learning techniques outperformed human-generated baselines. They also discovered that three traditional machine learning methods, Naive Bayes, Maximum Entropy Classification, and Support Vector Machines, performed worse than topic-based classification in sentiment classification tasks. For nonlinear modeling tasks such as text data classification, traditional machine learning struggles to handle massive data. With the development of neural networks, many text mining tasks have been conducted based on Long Short-Term Memory (LSTM) recurrent neural networks. Duyu Tang (2015) and others developed two target-dependent LSTM models [5], and empirical results showed that integrating target information into LSTM significantly improved classification accuracy. LSTM has been widely used for text sentiment classification for a long time. Although LSTM introduced long-term memory to address the vanishing gradient problem, it still struggles with long texts (e.g., those exceeding 500 words). With the introduction of attention mechanisms [6], Transformer structures have become the best choice for handling long sequential data. Recently, the state-of-the-art (SOTA) model is represented by the BERT model, and Zhengjie Gao (2019) developed a TD-BERT model [7], which achieved the best performance on text classification tasks across multiple datasets. Furthermore, the competition landscape in natural language processing has been disrupted by large-scale language models (LLM), with generative language models led by GPT (Generative Pre-Training) breaking multiple records in natural language processing and potentially becoming the next major research focus in the field of artificial intelligence.

In terms of sentiment classification research based on stock markets, in international research, Liang-Chih Yu and others proposed a stock news sentiment binary classification model using specific emotion-triggering words [8], and they demonstrated that their proposed method outperformed the previously proposed extension methods based on pointwise mutual information (PMI). However, specific trigger words rely heavily on the quality and level of manual annotations. Subsequently, Enric Junqué de Fortuny and others constructed a corpus using all articles from the online version of Flemish newspapers from 2007 to 2012, and they annotated the sentiment polarity of the corpus using the Pattern module in Python. Finally, they combined traditional technical indicators with sentiment indicators to conduct regression predictions on several stocks in the Brussels Euronext Exchange. The results showed that combining sentiment indicators with technical indicators performed better in terms of returns and Sharpe ratio. The aforementioned literature mainly focused on binary sentiment classification, which still falls short compared to the numerous emotions observed in reality. Building upon previous research, Qian Li and others categorized market sentiment as fear, anger, joy, sadness, anger, and disgust. They found that emotion-related words had a certain correlation with the overall trend of the stock market. However, this study also discovered that market sentiment alone cannot achieve the expected level of predicting the stock market. To validate the contribution of emotional factors to existing multifactor models, Wojarnik and others compared the quality of confusion matrices. They verified that the introduction of sentiment indicators increased the predictive value of models using machine learning mechanisms and found that emotional factors played a significant role in both short-term arbitrage and long-term investment.

\section{Method}
\subsection{Data}
We follow the following process to launch our experiments. First, we build the BERT model, finetune it using relevant data, and then test our model on stock news data.
\subsubsection{Training Data}
Before performing sentiment annotation on the text related to Oriental Fortune, this paper first needs to fine-tune the pre-trained BERT model, so it must have data with sentiment polarity labels related to the task of this paper. However, using the data from Oriental Fortune Stock Bar is unrealistic, because it relies on manual annotation of tens of thousands of data, so we have to rely on existing labeled datasets, here we use THUCTC Chinese text dataset. THUCTC (THU Chinese Text Classification) is a Chinese text classification toolkit launched by the Natural Language Processing Laboratory of Tsinghua University, which uses historical data from Sina News from 2005 to 2011 to label, and is divided into 14 categories according to the classification of Sina Weibo. This paper selects six sections: finance, stock, society, technology, politics, and entertainment as the training data for the model. A total of 481732 news data, of which there are 259489 positive samples and 222243 negative samples.

Before inputting the data into the BERT model, we need to do some preprocessing, including word segmentation, conversion to BERT model input format, etc. In this process, we use jieba, a Chinese word segmentation tool. jieba adopts a word segmentation method based on prefix dictionary, that is, splitting the text to be segmented into a series of vocabulary segments. At the same time, a stop word list is introduced to exclude some meaningless words. Subsequently, the data needs to be converted into the input format of BERT, that is, each text is converted into a vector, which contains information such as word vectors and position vectors. At the same time, in order to adapt to the MLM training mode of BERT, we need to randomly add [MASK] tags in the input vector for model training. The following figure summarizes the distribution of data and the pre-training process.

\subsubsection{Test Data}
The text data of this study comes from The East Money Stock BBS (http://guba.eastmoney.com/list,300059.html) and uses Python’s Request library and Xpath function to crawl elements such as comment title, popularity, time, etc. Finally, about 90,000 comment titles from January 1, 2019 to January 1, 2020 are selected as experimental data.

First of all, in order to better measure the effects of different sentiment analysis methods on this dataset, 500 data are randomly extracted from the dataset for manual annotation. It is agreed that 0 is negative comment and 1 is positive comment. Finally, we got 149 positive comments and 351 negative comments.

Among the 149 positive comments, “rise” was mentioned the most, 15 times, followed by “buy”, mentioned 9 times. Among the 351 negative comments, “sell” was mentioned the most, 19 times, followed by “fall”, mentioned 17 times.

From the above, it can be seen that both positive and negative comments have a clear view and prediction of the stock price trend. Analyzing these data can help investors better understand the market situation and the views of other investors, and thus influence the real stock price through the consensus mechanism. It is worth noting that the stock market is not solely affected by emotions, so we need to combine other stock price factors to develop our automatic trading model. First, we will use pre-training and fine-tuning BERT to automatically extract and classify positive and negative sentiments.

\subsection{BERT model}

n the fine-tuning stage, the BERT model will be used for specific NLP tasks. In this stage, the output of the pre-trained model needs to be connected with the task-related output and then fine-tuned. In the fine-tuning stage, different fine-tuning techniques can be used, such as multilayer perceptron (MLP) layer or convolutional neural network (CNN) layer as classifiers, to better adapt to specific NLP tasks. In the fine-tuning stage, the model is trained with labeled data and evaluated with evaluation data sets. If the model performs well, it can be used for practical applications. It is worth noting that in the fine-tuning stage, the hyperparameters of the pre-trained model can be adjusted, such as learning rate, batch size and number of training rounds, to optimize model performance. A technique called Masked Language Model (MLM) is used in this stage, which masks some words in the input text (represented by “[MASK]”) and then lets the model predict these masked words. This allows the model to learn contextual information, because in training, the model can only rely on known words to predict the words at masked positions. The pre-training objective is to minimize the prediction error of the masked words and the next sentence in the input text, so that massive data can be trained without manual annotation.

In order to adapt to Chinese sentiment analysis tasks, this paper uses bert-base-chinese pre-trained model. This model is based on 21128 words for MLM training as described above, and has been proven to be a well-performing pre-trained model by a large number of downstream tasks.

\section{Experiment and Result}
\subsection{Evaluation index}

Since the task of this article is a classification task, we use AUC as an indicator to evaluate the model.

AUC (Area Under the ROC Curve) is an indicator that measures the predictive performance of a binary classification model, representing the area under the ROC curve. In the ROC curve, the true positive rate (TPR) is the y-axis and the false positive rate (FPR) is the x-axis. We can use this curve to evaluate the prediction results of a binary classification model. The closer the ROC curve is to the upper left corner, the better the model’s performance, and its true positive rate is higher than its false positive rate. The AUC indicator is the area under the ROC curve, and the closer its value is to 1, the better the model’s classification effect. When the value is equal to 0.5, it is equivalent to random classification.

\subsection{Training Result}
This paper is based on the GPU server of the AutoDL platform for training, using an RTX 3090, and based on Python3.8 and Pytorch1.11.0 to build the model. The following figure shows the comparison between the trained model and the baseline model, showing that fine-tuning plays a relatively large role in improving classification accuracy.

\begin{figure}[htbp!]
	\centering
	\includegraphics[width=1.0\linewidth]{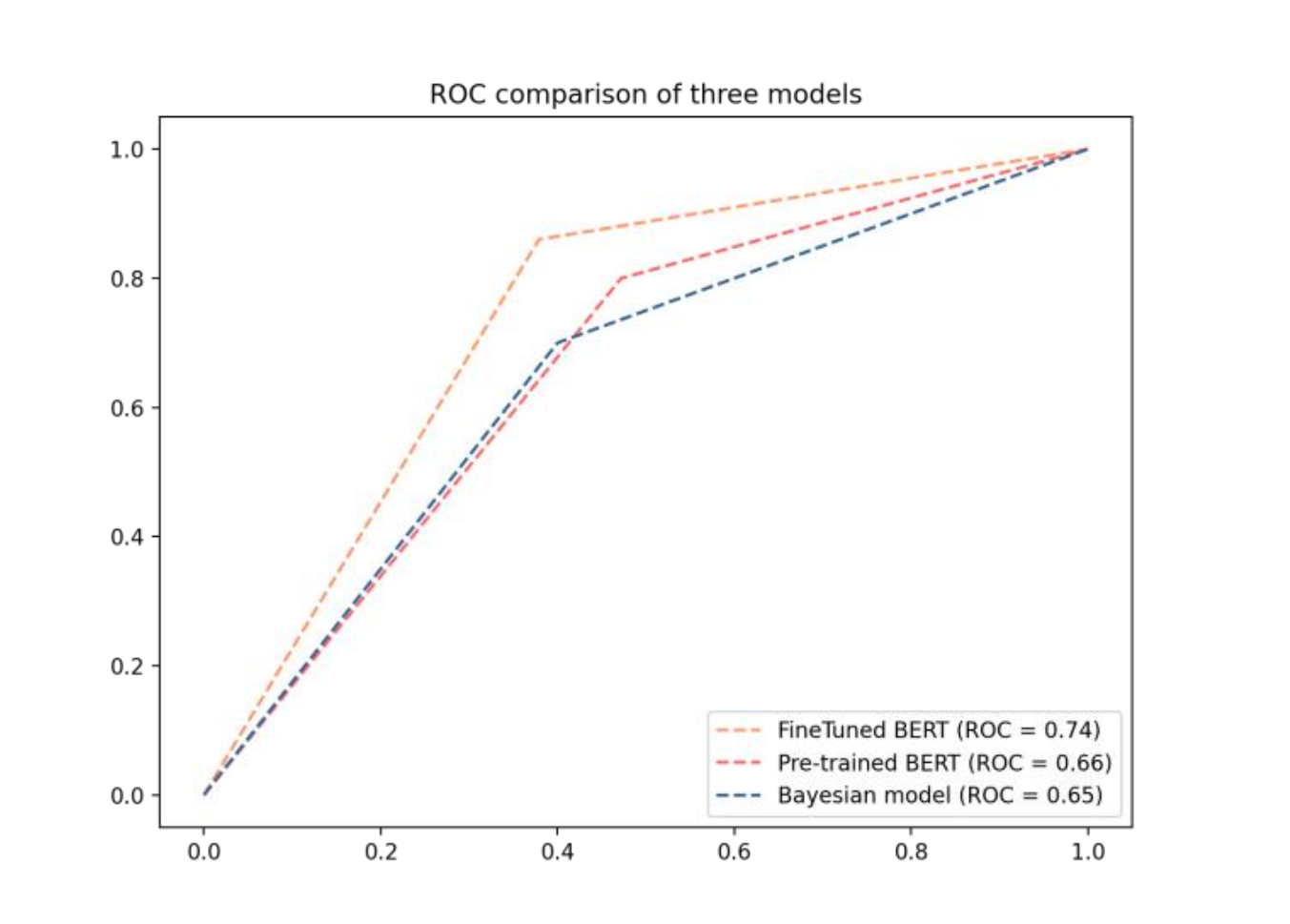}
	\caption{Classification Result}
	\label{datasplit}
\end{figure}

According to the above experimental results, we found that the sentiment classification performance based on the BERT model is better than the traditional Bayesian classification model. In the training process, the average loss function value of each round gradually decreases, and after five rounds of training, it basically converges, indicating that the model has tended to be stable. And the model has good AUC performance on the test set and validation set, indicating that the model has high accuracy and robustness in sentiment classification tasks.

The Figure 2 shows the trend of the change rate and sentiment mean value of the first 100 trading days. It can be found that there is a certain correlation between the two trends. At the same time, through correlation test, the absolute value of the correlation coefficient between the stock price change rate sequence and the sentiment value sequence is 0.17. As a reference, the average absolute value of the correlation coefficient of 191 factors in Guotai Junan 191 factor library relative to the price sequence mentioned in the following text is 0.1233, which shows that the factor we mined through sentiment analysis can have a certain explanatory ability for stock price trends.

\begin{figure*}[htbp!]
	\centering
	\includegraphics[width=1.0\linewidth]{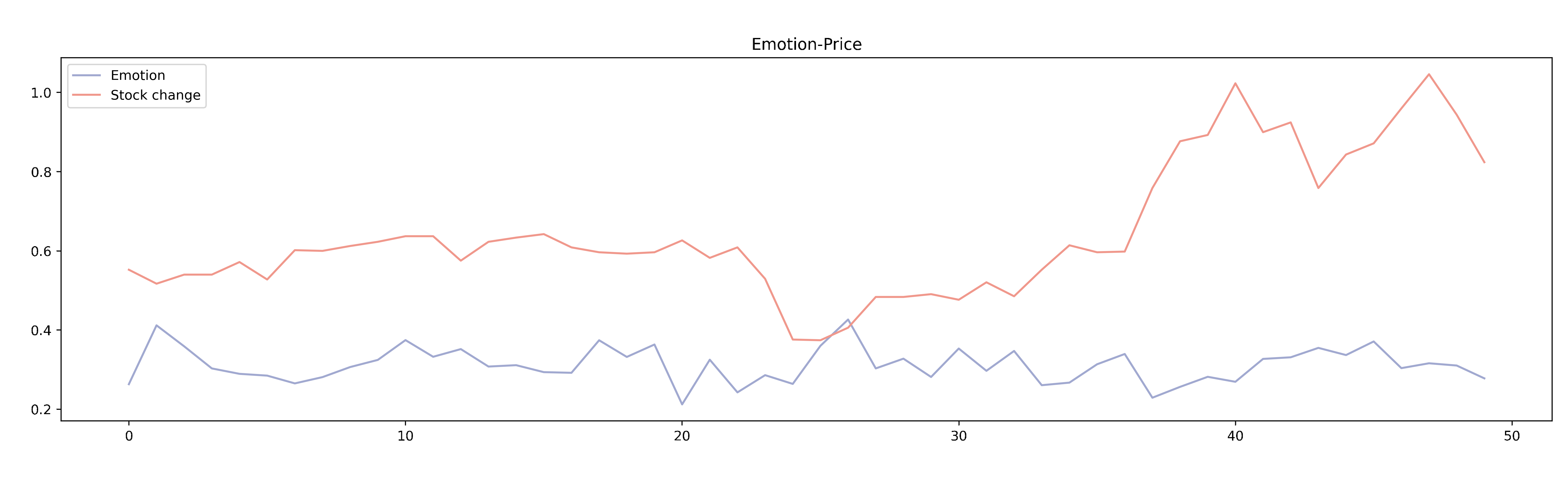}
	\caption{Emotion-Price}
	\label{datasplit}
\end{figure*}

Therefore, we found that fine-tuning the BERT model with relevant datasets makes it perform better in dealing with the relative tasks. This shows that fine-tuning plays a decisive role in transferring from a general model to a specific task. In summary, the BERT model can perform sentiment classification on the comment information of Oriental Fortune Stock Bar relatively well, which helps us to develop signal factors using sentiment classification information in the next step, and test its return rate and other strategy indicators through backtesting.

\section{Auto-Trade Test}
\subsection{Experiment}

The basic factors of this paper are based on the Guotai Junan 191 factor library (alpha 101), which is a comprehensive stock investment factor library developed by Guotai Junan Securities Company. The factor library contains 191 investment factors, covering the market, industry, and company levels, as well as financial, valuation, technical and other aspects. Among them, value factors mainly reflect the valuation level of stocks, such as price-to-earnings ratio, price-to-book ratio, etc.; growth factors focus on the future growth potential of enterprises, such as EPS growth rate, operating income growth rate, etc.; quality factors consider the profitability, debt repayment ability, operational ability and other aspects of enterprises; momentum factors reflect the trend and direction of stock prices; volatility factors measure the degree of fluctuation of stock prices; liquidity factors consider the liquidity and trading volume of stocks and other factors.

The prediction indicator of this paper is the average rise and fall rate of the next five trading days. And in order to better explain the impact of factors on the dependent variable, this paper uses linear regression to construct a signal model.

This paper built an automatic trading evaluation system based on Python, which can read a set of market price data and action list, automatically perform backtesting verification and output relevant charts and evaluation indicators. After multiple rounds of testing, the final buy and sell strategy selected by this paper is: buy 1000 shares of the target when the prediction is greater than the 90th percentile of the test set, and sell 500 shares of the target when it is less than the 10th percentile. At the same time, in order to avoid frequent trading, we also stipulate that each purchase must be held for at least 10 days before making the next decision, to avoid decision errors caused by market fluctuations and excessive transaction fees.

We set the transaction fee to 0.5\%, and the initial capital to 100000. We tested the alpha191 strategy (hereinafter referred to as alpha191), the strategy that uses only the emotion factor (hereinafter referred to as Emotion), and the strategy that combines the two (hereinafter referred to as alpha191+Emotion). At the same time, we tested the strategy of buying stocks on the first day and holding them all the time as a benchmark (hereinafter referred to as buy-and-hold). The final test results are shown in the Figure 3.

\begin{figure*}[htbp!]
	\centering
	\includegraphics[width=1.0\linewidth]{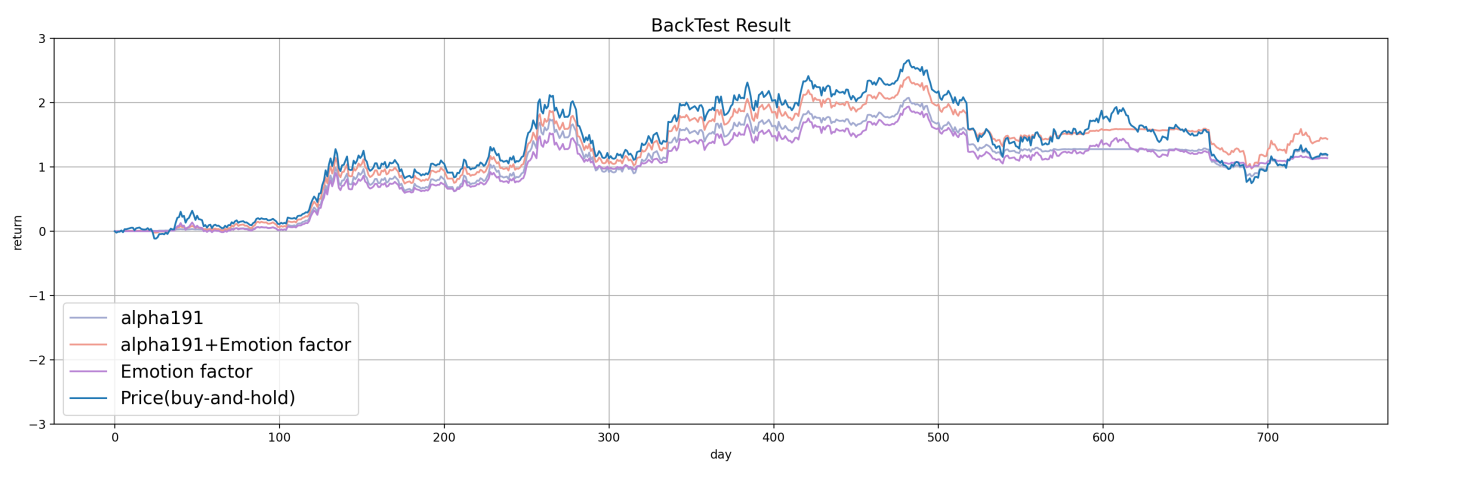}
	\caption{Backtest Result}
	\label{datasplit}
\end{figure*}

We can see from the figure that the return obtained by using the Emotion strategy alone did not exceed the benchmark, and the return obtained by using the alpha191 strategy alone was not much different from the benchmark return. Combining the Emotion strategy with the alpha191 strategy can achieve higher excess returns. The evaluation indicators of the four strategies are shown in the following table:

\begin{table*}[htbp!]
\begin{center}

\begin{tabular}{@{}
>{\columncolor[HTML]{FFFFFF}}c 
>{\columncolor[HTML]{FFFFFF}}c 
>{\columncolor[HTML]{FFFFFF}}c 
>{\columncolor[HTML]{FFFFFF}}c 
>{\columncolor[HTML]{FFFFFF}}c @{}}
\toprule
                 & \textbf{buy-and-hold} & \textbf{alpha191} & \textbf{Emotion} & \textbf{Emotion+alpha191} \\ \midrule
Maximum Profit   & 166.14\%              & 207.53\%          & 193.99\%         & 239.94\%                  \\
Maximum Drawdown & 191.47\%              & 122.35\%          & 95.32\%          & 142.67\%                  \\
Return           & 118.98\%              & 117.58\%          & 113.58\%         & 143.31\%                  \\
Excess Return    & \textbackslash{}      & -1.40\%           & -4.00\%          & 29.73\%                   \\ \bottomrule
\end{tabular}
\caption{Backtest Result}
\end{center}
\end{table*}

It can be seen from the table that the Emotion+alpha191 strategy performs the best in the three indicators of maximum return, final return, and excess return. In terms of drawdown control, the Emotion strategy performs the best. At the same time, only the Emotion+alpha191 strategy shows a positive excess return, beating the target’s own increase.

\subsection{Analyze}

The above results show that using the emotion factor as a separate investment strategy can bring a higher return rate than the benchmark buy-and-hold strategy. However, this return rate still does not exceed the return rate obtained by using the alpha191 factor model. This indicates that although the emotion factor is an effective factor, other factors still need to be considered in investment decisions. At the same time, we found that combining the emotion factor with the alpha191 factor model can further increase the return rate. This proves that the emotion factor can be a good complement to the traditional factor model, which can help investors better understand the market situation and formulate more effective investment strategies. In future investment decisions, investors can consider the emotion factor as one of the important investment indicators, to better avoid risks and increase returns.

\section{Conclusion}

The main work of this paper is based on the post title data from the East Money Stock BBS, using natural language processing technology and machine learning algorithms, to develop a sentiment analysis model and apply it to quantitative investment. Specifically, this paper uses the BERT model to perform sentiment analysis on the post titles, and uses the sentiment polarity values as trading signal factors, and develops an automatic trading system. The experimental results show that the trading factors extracted by the sentiment analysis method based on the BERT model can be a good complement to the existing factor system, proving that incorporating the emotional information of market participants into stock trading decisions is a feasible way.

This paper uses advanced natural language processing technologies and machine learning algorithms such as BERT to deeply mine and analyze the emotional information in the stock market. The application of these technologies makes the sentiment analysis model have higher accuracy and robustness. At the same time, this paper develops an automatic trading system based on sentiment analysis signal factors. This general system has a convenient and easy-to-use interface and calling method, which can be easily used for subsequent research. At the same time, this paper does not stop at the emotional classification of text, but verifies the effectiveness of the sentiment analysis model in the stock market through experiments, and performs backtesting verification on the automatic trading system. The experimental results prove that the inclusion of emotional factors in the regression equation improves the return rate and model explanatory power. These experimental results provide important reference for subsequent research.

However, due to objective reasons, this paper also has some shortcomings. The training data set used in this paper comes from the THUCTC Chinese text classification toolkit. Although this data set is large in scale, it only contains news data, not real-time data in the stock market. Therefore, this model may have some limitations in practical applications. Secondly, although this paper uses advanced natural language processing technologies and machine learning algorithms such as BERT, there is still room for improvement in the accuracy of the sentiment analysis model. Future research can explore more refined feature engineering and model optimization methods. In addition, the automatic trading system strategy based on sentiment analysis signal factors is temporarily simple in this paper, and does not consider position limits, stop-loss and stop-profit situations in detail. Future research needs to explore more deeply how to effectively manage trading risks and improve the stability and reliability of the trading system.

\end{document}